\documentclass[12pt]{article}

\begin{document}
\begin{titlepage}
\title{Axi-Dilaton Gravity in $D  \geq 4$
Dimensional Space-times with Torsion}

\author{H. Cebeci\footnote{E.mail: hcebeci@metu.edu.tr}\\ {\small
Department of Physics, Middle East Technical
University,}\\{\small 06531 Ankara,
Turkey}\\  \\ T. Dereli\footnote{E.mail: tdereli@ku.edu.tr}\\
{\small  Department of Physics, Ko\c{c}
University,}\\{\small 034450 \.{I}stanbul, Turkey}}
\date{ }

\maketitle

\bigskip

\begin{abstract}

\noindent We study models of axi-dilaton gravity in space-time
geometries  with torsion. We discuss conformal rescaling rules in
both Riemannian and non-Riemannian formulations. We give static,
spherically symmetric solutions and examine their singularity
behaviour.

\end{abstract}

\bigskip


\end{titlepage}

\section{Introduction}

Gravitational interactions are formulated on a space-time manifold
$M$ equipped with a metric tensor field $g$ and a metric
compatible connection $ \nabla $ defined on the bundle of
orthonormal frames. Most commonly, interactions coupled with
gravity are studied in a geometry where the connection $ \nabla $
is constrained to be the unique torsion-free Levi-Civita
connection. In this context, massive test particles are postulated
to follow time-like geodesics associated with space-time metric
and torsion-free connection. On the other hand a metric compatible
connection with torsion provides new independent degrees of
freedom. It has been shown that the scalar field interactions
coupled with gravity can yield connections with non-zero torsion
\cite{dereli1}. In that case, space-time history of particles may
be determined by the autoparallels of a connection with torsion
\cite{dereli3,dereli4,dereli8}. We know that the independent
variation of any action with respect to connection  determines
space-time torsion. In particular, bosonic part of effective
superstring interactions can produce a torsion that is
proportional to the gradient of the dilaton (scalar) field. Hence,
it would be of interest to formulate such type of interactions in
frames where torsion exists.

It is an exciting conjecture that all superstring models belong to
an eleven dimensional M-theory that accommodates their apparent
dualities. M-theory as a classical theory can be considered in a
low-energy limit where only the low-lying excitation modes
contribute to an effective field theory. As such it would be the
same as $D=11$ dimensional supergravity theory. A subsequent
Kaluza-Klein reduction to $D=10$ dimensions would bring it to a
string model whose gravitational sector consists of space-time
metric tensor $g$, dilaton scalar $\phi $ and the axion potential
$(p+1)$-form $A$ that would minimally couple to $p$-branes. We
call such an effective gravitational field theory an axi-dilaton
gravity in $D$ dimensions. Axi-dilaton gravity theory can be
studied in the Einstein frame. However, by working out the theory
in Brans-Dicke frame \cite{brans}, one can see the difference
between formulation of theory with a torsion-free connection and
formulation with a connection with torsion. In the latter case, we
vary the action treating the metric and the connection as
independent variables. We have shown that the corresponding field
equations in both cases with or without torsion are equivalent
provided a shift in the Brans-Dicke coupling parameter $\omega$ is
allowed. We further assume a direct coupling of the $k^{th}$ power
of the dilaton scalar with the axionic kinetic term. The conformal
scaling properties are examined in both geometries. In Section:3
we investigate a class of static, spherically symmetric solutions
which depend on the coupling parameters $\omega$ and $k$ in
dimensions $D \geq 4$. In particular, we point out a class of
conformal black hole solutions obtained for the scale invariant
parameter values.

\section{Axi-dilaton Gravity In $D$ Dimensions}

We start with an action
\begin{equation}
I [ g, \phi, A ] = \int_{M} {\cal{L}}
\end{equation}
where the Lagrangian density $D$-form $\cal{L}$ is given in
Brans-Dicke frame in a geometry based on Riemannian formulation,
by imposing as a constraint that the connection is Levi-Civita:
\begin{equation}
{\cal L}=\frac{\phi}{2}\, R^{ab}\wedge \ast \left( e_{a}\wedge
e_{b}\right) - \frac{\omega}{2 \phi} d\phi \wedge \ast d\phi
-\frac{\phi ^{k}}{2}H\wedge \ast H . \label{1}
\end{equation}
Here the basic gravitational variables are the co-frame 1-forms
$e^{a}$ in terms of which the space-time metric $g=\eta
_{ab}e^{a}\otimes e^{b}$ where $\eta _{ab}=diago(-+++++...)$.
Hodge $\ast$-map is defined so that the oriented volume form $\ast
1=e^{0}\wedge e^{1}\wedge ...\wedge e^{D-1}$. Levi-Civita
connection 1-forms $^{(0)}\omega^{a}\,_{b}$ are obtained from the
first Cartan structure equations
\begin{equation}
de^{a}+^{(0)}\omega^{a}\,_{b}\wedge e^{b}=0
\label{2}
\end{equation}
where the metric compatibility requires $^{(0)} \omega_{ab}= -
^{(0)} \omega_{ba}$ and corresponding curvature 2-forms are
obtained from the second Cartan structure equations
\begin{equation}
^{(0)}R^{ab}=d\, ^{(0)}\omega ^{ab}+^{(0)}\omega^{a}\,_{c}\wedge
^{(0)}\omega ^{cb} .
\label{3}
\end{equation}
$\phi $ is the dilaton 0-form and $H$ is a $(p+2)$-form field that
is derived from the axion potential $(p+1)$-form  $A$ so that
$H=dA$. $\omega$ and $k$ are real coupling parameters. Co-frame
$e^{a}$ variations of this action lead to the Einstein field
equations
\begin{equation}
\frac{1}{2}\phi\,\, ^{(0)}R^{ab}\wedge \ast (e_{a}\wedge e_{b}\wedge
e_{c})=-\frac{\omega }{2 \phi} \tau_{c}[\phi]-\frac{\phi^{k}}{2}
\tau_{c}[H]-^{(0)}D(\iota _{c}(\ast d\phi ))
\label{4}
\end{equation}
where dilaton and axion stress-energy $(D-1)$-forms are given,
respectively, by
\begin{equation}
\tau_{c}[\phi]=(\iota_{c}d\phi \wedge \ast d\phi + d\phi \wedge
\iota_{c}(\ast d\phi))
\label{5}
\end{equation}
and
\begin{equation}
\tau_{c}[H]=(\iota_{c}H \wedge \ast H - (-1)^{p} H \wedge \iota
_{c}(\ast H) ) .
\label{6}
\end{equation}
$\phi$ variation of (\ref{1}) yields
\begin{equation}
\frac{1}{2} \,^{(0)}R^{ab} \wedge \ast(e_{a} \wedge e_{b})=-\omega d
\left(\frac{\ast d\phi}{\phi} \right)-\frac{\omega}{2 \phi^{2}} d\phi \wedge \ast d\phi + k \frac{\phi^{k-1}}{2} H \wedge \ast H
\label{7}
\end{equation}
We trace (\ref{4}) by considering its exterior multiplication by
$e^{c}$ and multiply (\ref{7}) by $(D-2) \phi$. The resulting two
equations are then subtracted side by side to obtain the dilaton
field equation
\begin{equation}
\left( \omega +\frac{D-1}{D-2}\right) d \ast d\phi =\frac{\phi
^{k}}{2}\alpha H\wedge \ast H  , \label{8}
\end{equation}
where $\alpha = \frac{2p-(D-4)}{D-2} + k$. Finally, independent
axion potential $A$ variations lead to
\begin{equation}
d(\phi ^{k}\ast H)=0
\label{9}
\end{equation}
with $dH=0$.

Next we consider the following action in which connection 1-forms
are varied independently of the metric of space-time:
\begin{equation}
{\cal L}=\frac{\phi}{2} R^{ab}\wedge \ast \left( e_{a}\wedge
e_{b}\right) - \frac{c}{2\phi } d\phi \wedge \ast d\phi
-\frac{\phi ^{k}}{2} H \wedge \ast H . \label{10}
\end{equation}
Co-frame variations of this action give the Einstein field
equations
\begin{equation}
\frac{1}{2}\phi R^{ab}\wedge (e_{a}\wedge e_{b}\wedge
e_{c})=-\frac{c}{2\phi } \tau_{c} [\phi] - \frac{1}{2} \phi ^{k}
\tau_{c} [H] . \label{11}
\end{equation}
with $\tau_{c} [\phi]$ and $\tau_{c} [H]$ are as given by
(\ref{5}) and (\ref{6}), respectively. Scalar field variations of
the action give
\begin{equation}
\frac{1}{2} R^{ab} \wedge (e_{a} \wedge e_{b})=-c d \left(\frac{\ast d
\phi}{\phi} \right)-\frac{c}{2 \phi^{2}} d\phi \wedge \ast d\phi + k
\frac{\phi^{k-1}}{2} H \wedge \ast H .
\label{12}
\end{equation}
When we trace (\ref{11}) and compare it with  (\ref{12})
multiplied by $(D-2) \phi$, we obtain the dilaton field equation
\begin{equation}
c\,d\ast d\phi =\frac{\alpha }{2}\phi ^{k} H \wedge \ast H .
\label{13}
\end{equation}
Independent connection variations of (\ref{10}) lead to
\begin{equation}
D \left( \frac{\phi }{2}\ast (e^{a}\wedge e^{b}) \right)=0
\label{14}
\end{equation}
from which we can readily solve for the torsion 2-forms:
\begin{equation}
T^{a}=e^{a}\wedge \frac{d\phi }{(D-2)\phi } . \label{15}
\end{equation}
 We can decompose the connection 1-forms as
\begin{equation}
\omega^{a}\,_{b}=^{0}\omega^{a}\,_{b}+K^{a}\,_{b}
\label{16}
\end{equation}
where contorsion one-forms $K^{a}\,_{b}$ satisfy
\begin{equation}
K^{a}\,_{b}\wedge e^{b}=T^{a} .
\label{17}
\end{equation}
Substitution of (\ref{15}) into (\ref{17}) gives
\begin{equation}
K^{a}\,_{b}=\frac{1}{(D-2)\phi }(e^{a}\iota _{b}d\phi - e_{b}\iota
^{a}d\phi ) .
\label{18}
\end{equation}
Curvature 2-forms $R^{ab}$ can be similarly decomposed as
\begin{equation}
R^{ab}=^{(0)}R^{ab}+^{(0)}DK^{ab}+K^{a}\,_{c}\wedge K^{cb}
\label{19}
\end{equation}
where
\begin{equation}
^{(0)}DK^{ab}=dK^{ab}+^{(0)}\omega^{b}\,_{c} \wedge
K^{ac}+^{(0)}\omega^{a}\,_{c} \wedge K^{cb} .
\label{20}
\end{equation}
Then we calculate
\begin{eqnarray}
R^{ab} \wedge \ast ( e_{a} \wedge e_{b} \wedge e_{c} ) = ^{(0)}R^{ab}
\wedge
\ast (e_{a} \wedge e_{b} \wedge e_{c} ) +\frac{2}{\phi}\, ^{(0)}D(\iota_{c}
(\ast d\phi )) \nonumber \\
 - \frac{2(D-1)}{(D-2) \phi^{2}} d\phi \wedge \iota_{c} (\ast d\phi) -
\frac{D-1}{(D-2)\phi^{2}} \iota_{c} (d\phi \wedge \ast d\phi )
\label{21}
\end{eqnarray}
and
\begin{equation}
R^{ab} \wedge \ast (e_{a} \wedge e_{b} ) = ^{(0)}R^{ab} \wedge \ast (
e_{a} \wedge e_{b} )-\frac{2(D-1)}{(D-2)}d \left(\ast \frac{d\phi}{\phi}
\right) - \frac{D-1}{(D-2) \phi^{2}} d \phi \wedge \ast d \phi .
\label{22}
\end{equation}
If we insert (\ref{22}) into (\ref{10}), action density reduces to
\begin{equation}
{\cal L}=\frac{1}{2}\phi\,\, ^{(0)}R^{ab}\wedge \ast (e_{a}\wedge
e_{b}) - \left(c-\frac{D-1}{D-2} \right)
\frac{1}{2\phi }d\phi \wedge \ast d\phi -\frac{\phi ^{k}}{2}H\wedge \ast H
\label{23}
\end{equation}
upto a closed form. Substituting (\ref{21}) into the Einstein
field equations (\ref{11}), we obtain
\begin{eqnarray}
\frac{1}{2}\phi\, ^{(0)}R^{ab}\wedge \ast (e_{a}\wedge e_{b}\wedge
e_{c})&=& - \left(c- \frac{D-1}{D-2}\right) \frac{1}{2\phi } \tau_{c} [ \phi ] \nonumber \\
& & - \frac{ \phi^{k} }{2} \tau_{c} [H] - ^{0}D \left( \iota_{c}( \ast d\phi ) \right) .
\label{24}
\end{eqnarray}
Similarly, substituting (\ref{22}) into the dilaton field equation
(\ref{12}), we obtain
\begin{eqnarray}
\frac{1}{2}\, ^{(0)}R^{ab} \wedge \ast ( e_{a} \wedge e_{b} ) &=&
\frac{\left(c-\frac{D-1}{D-2} \right)}{2 \phi^{2}} d \phi \wedge \ast d
\phi - \left(c-\frac{D-1}{D-2} \right) \frac{1}{\phi} d ( \ast d \phi)
\nonumber \\
& & + k \frac{\phi^{k-1}}{2} H \wedge \ast H .
\label{25}
\end{eqnarray}
We have thus shown that if the coupling constants are identified
as
\begin{equation}
\omega = c - \frac{D-1}{(D-2)} ,
\label{26}
\end{equation}
the field equations (\ref{24}) and (\ref{25}) are equivalent to
the field equations (\ref{4}) and (\ref{7}).

Let us now consider conformal rescalings of the metric induced by
the co-frame rescalings
\begin{equation}
e^{a} \rightarrow e^{\sigma (x)} e^{a} .
\label{27}
\end{equation}
These imply the transformation
\begin{equation}
^{(0)} \omega_{ab} \rightarrow ^{(0)} \omega_{ab} - e_{b} \iota_{a} d \sigma + \iota_{b} d \sigma e_{a}
\label{28}
\end{equation}
of the Levi-Civita connection 1-forms. If we also postulate the following rescaling
of the Brans-Dicke scalar field
\begin{equation}
\phi \rightarrow e^{-(D-2) \sigma } \phi ,
\label{29}
\end{equation}
then a straightforward calculation shows that the action (\ref{1})
is scale invariant for $ \omega = - \frac{D-1}{D-2} $ and $ k = -
\frac{2p+4-D}{D-2} $ , or for $ c=0$ and $ \alpha = 0 $. In terms
of the geometry described by the action (\ref{10}), the above
rescaling rules imply the transformation
\begin{equation}
K_{ab} \rightarrow K_{ab} + \iota_{a} d \sigma e_{b} - \iota_{b} d \sigma e_{a}
\label{30}
\end{equation}
so that the connection with torsion does not scale:
\begin{equation}
\omega_{ab} \rightarrow \omega_{ab} .
\label{31}
\end{equation}
Hence
\begin{equation}
R_{ab} \rightarrow R_{ab}
\label{32}
\end{equation}
and
\begin{equation}
T^{a} \rightarrow e^{ \sigma } ( T^{a} + d \sigma \wedge e^{a} ) .
\label{33}
\end{equation}

We can reformulate our axi-dilaton gravity in the so-called
Einstein frame by adopting the co-frames
\begin{equation}
{\tilde{e}}^{a}=\left(\frac{\phi}{\phi_{0}}\right)^{\frac{1}{(n-1)}}
e^{a}
\label{34}
\end{equation}
where $\phi_{0}$ is a constant. The new co-frames
${\tilde{e}}^{a}$ become orthonormal with respect to space-time
metric
\begin{equation}
\tilde{g}=\left(\frac{\phi}{\phi_{0}}\right)^{\frac{2}{(n-1)}} g .
\label{35}
\end{equation}
In terms of this metric the associated Hodge dual is denoted by
$\tilde{\ast}$. For an arbitrary frame independent $p$-form
$\Omega$,
\begin{equation}
\ast \Omega = \left( \frac{\phi}{\phi_{0}}
\right)^{\frac{2p-(n+1)}{(n-1)}} \tilde{\ast} \Omega
\label{36}
\end{equation}
In the reformulation of action (\ref{1}) in terms of $\tilde{g}$,
new connection fields ${\tilde{\omega}}^{ab}$ can be written in
terms of $ ^{(0)} \omega^{ab} $ as
\begin{equation}
{\tilde{\omega}}^{ab} = \Gamma^{ab} + ^{(0)} \omega^{ab}
\label{37}
\end{equation}
where,
\begin{equation}
\Gamma^{ab} = \frac{1}{(D-2) \phi}(e^{a} \iota^{b} d \phi - e^{b}
\iota^{a} d \phi ) .
\label{38}
\end{equation}
The corresponding  curvature 2-forms become
\begin{equation}
\tilde{R}^{ab} = ^{(0)} R^{ab} + ^{(0)} D \Gamma^{ab} + \Gamma^{ac} \wedge
\Gamma_{c}\,^{b} .
\label{39}
\end{equation}
In terms of $\tilde{g}$, (\ref{1}) becomes
\begin{equation}
{\cal L} = \frac{1}{2} \phi_{0} \tilde{R}^{ab} \wedge \tilde{\ast}
( {\tilde{e}}_{a} \wedge {\tilde{e}}_{b} ) - \frac{c}{2} \phi_{0}
\frac{1}{\phi^{2}} d \phi \wedge \tilde{\ast} d \phi -
\frac{\phi^{\alpha}}{2} \phi_{0}^{(k - \alpha )} H \wedge
\tilde{\ast} H , \label{40}
\end{equation}
upto a closed form. Introducing a massless scalar field $\Phi= \ln
\mathopen| \frac{\phi}{\phi_{0}} \mathclose|$, (\ref{40}) reads
\begin{equation}
{\cal L} = \frac{1}{2} \phi_{0} \tilde{R}^{ab} \wedge \tilde{\ast}
( \tilde{e}_{a} \wedge \tilde{e}_{b} ) - \frac{c}{2} \phi_{0} d
\Phi \wedge \tilde{\ast} d \Phi - \frac{1}{2} (\phi_{0})^{k} \exp
( \alpha \Phi ) H \wedge \tilde{\ast} H . \label{41}
\end{equation}
Einstein field equations obtained by co-frame variations of
(\ref{41}) are
\begin{equation}
\frac{1}{2} \phi_{0} \tilde{R}^{ab} \wedge \tilde{\ast} ( \tilde{e}_{a} \wedge \tilde{e}_{b} \wedge \tilde{e}_{c} ) = - \frac{c}{2} \phi_{0} \tilde{\tau}_{c} [\Phi]- \frac{1}{2} (\phi_{0})^{k} e^{ \alpha \Phi } \tilde{\tau}_{c} [H] ,
\label{42}
\end{equation}
where
\begin{equation}
\tilde{\tau}_{c} [\Phi]= \{ \tilde{\iota}_{c} d \Phi \wedge \tilde{\ast} d \Phi + d \Phi \wedge \tilde{\iota}_{c} (\tilde{\ast} d \Phi ) \}
\label{43}
\end{equation}
and
\begin{equation}
\tilde{\tau}_{c} [H] = \{ \tilde{\iota}_{c} H \wedge \tilde{\ast} H - (-1)^{p} H \wedge \tilde{\iota}_{c} ( \tilde{\ast} H ) \} .
\label{44}
\end{equation}
On the other hand variations with respect to connection 1-forms
$\tilde{\omega}^{ab}$ yield
\begin{equation}
D ( \tilde{\omega} ) ( \tilde{\ast} ( \tilde{e}_{a} \wedge \tilde{e}_{b} ) ) = 0 ,
\label{45}
\end{equation}
from which we obtain $\tilde{T}^{a} = 0$. Finally, we give  the
scalar field equation
\begin{equation}
c \phi_{0} d (\tilde{\ast} d \Phi ) = \frac{1}{2} (\phi_{0})^{k} \alpha e^{ \alpha \Phi } H \wedge \tilde{\ast} H .
\label{46}
\end{equation}
and the axion field equation
\begin{equation}
 d \left( e^{ \alpha \Phi } \tilde{\ast} H \right) = 0 .
\label{47}
\end{equation}

Interestingly, by another conformal rescaling of the co-frames in
Einstein frame, we can obtain the so-called string frame action.
Applying the transformation
\begin{equation}
\hat{e}^{a} = \exp ( \frac{2 \Phi}{D-2} ) \tilde{e}^{a}
\label{48}
\end{equation}
where $\hat{e}^{a}$ are assumed to satisfy the torsion-free
structure equations
\begin{equation}
d \hat{e}^{a} + \hat{\omega}^{a}\,_{b} \wedge \hat{e}^{b} = 0 \, ,
\label{49}
\end{equation}
the action density (\ref{41}) becomes
\begin{eqnarray}
{\cal L} = e^{-2 \Phi} \left\{ \frac{1}{2} \phi_{0} \hat{R}^{ab}
\wedge \hat{\ast} ( \hat{e}_{a} \wedge \hat{e}_{b} ) - \frac{1}{2}
\phi_{0} \hat{k} d \Phi \wedge \hat{\ast} d \Phi \right\} -
\frac{1}{2} (\phi_{0})^{k} \exp ( \alpha_{0} \Phi ) H \wedge
\hat{\ast} H , \label{50}
\end{eqnarray}
upto a closed form where coupling parameters are redefined as
\begin{equation}
\alpha_{0} = ( 2 p + 4 - D ) \frac{3}{D-2} + k
\label{51}
\end{equation}
and
\begin{equation}
\hat{k} = c - \frac{4 (D-1)}{(D-2)} .
\label{52}
\end{equation}
Action density (\ref{50}) is called the string frame action in $D$
dimensions. We would like to remark that it is possible to start
directly from  (\ref{50}) and make independent co-frame
$\hat{e}^{a}$ and connection $\hat{\omega}^{ab}$ variations.
Independent connection variations yield
\begin{equation}
D (\hat{\omega}) (  e^{- 2 \Phi } \hat{\ast} ( \hat{e}_{a} \wedge \hat{e}_{b} ) ) = 0
\label{53}
\end{equation}
from which we can obtain torsion 2-forms $ \hat{T}^{a} =
\frac{2}{D-2} d \Phi \wedge \hat{e}^{a} $ \cite{dereli7}. Co-frame
variations on the other hand  yield
\begin{eqnarray}
\frac{1}{2} \phi_{0} e^{ - 2 \Phi } \hat{R}^{ab} \wedge \hat{\ast} ( \hat{e}_{a} \wedge \hat{e}_{b} \wedge \hat{e}_{c} ) &=& - \frac{1}{2} \phi_{0} \hat{k} e^{ - 2 \Phi } \hat{\tau}_{c} [\Phi] \nonumber \\
& & - \frac{1}{2} (\phi_{0})^{k} e^{ \alpha_{0} \Phi } \hat{\tau}_{c} [H] ,
\label{54}
\end{eqnarray}
where
\begin{equation}
\hat{\tau}_{c} [\Phi] = \{ \hat{\iota}_{c} d \Phi \wedge \hat{\ast} d \Phi + d \Phi \wedge \hat{\iota}_{c} ( \hat{\ast} d \Phi ) \}
\end{equation}
and
\begin{equation}
\hat{\tau}_{c} [H] = \{ \hat{\iota}_{c} H \wedge \hat{\ast} H - (-1)^{p} H \wedge \hat{\iota}_{c} ( \hat{\ast} H ) \} .
\end{equation}
The scalar field $\Phi$ variation of (\ref{50}) gives
\begin{eqnarray}
\phi_{0} e^{ - 2 \Phi } \hat{R}^{ab} \wedge \hat{\ast} ( \hat{e}_{a} \wedge \hat{e}_{b} ) = \phi_{0} \hat{k} e^{ - 2 \Phi } d \Phi \wedge \hat{\ast} d \Phi \nonumber \\
+ \hat{k} \phi_{0} d \left( e^{ - 2 \Phi } \hat{\ast} d \Phi \right) - \frac{1}{2} (\phi_{0})^{k} \alpha_{0} e^{ \alpha_{0} \Phi } H \wedge \hat{\ast} H .
\label{55}
\end{eqnarray}
We consider exterior multiplication of (\ref{54}) by $\hat{e}^{c}$ and then multiply the equation by $\frac{2}{2-D}$. If we subtract the resulting equation from (\ref{55}) and use (\ref{51}), we obtain the scalar field equation
\begin{equation}
\phi_{0} \hat{k} d \left( e^{ - 2 \Phi } \hat{\ast} d \Phi \right) = \frac{1}{2} (\phi_{0})^{k} \alpha e^{ \alpha_{0} \Phi } H \wedge \hat{\ast} H .
\end{equation}
Finally, the gauge field $A$ variation yields
\begin{equation}
d \left( e^{ \alpha_{0} \Phi } \hat{\ast} H \right) = 0 .
\end{equation}
The field equations without torsion in the string frame can be
determined exactly in the same way we explained above.

\section{Static, Spherically Symmetric Solutions}

In this section we investigate a class of static, spherically
symmetric solutions of the axi-dilaton field equations with
$p=D-4$, in the Brans-Dicke frame. Such solutions were studied
previously in the Einstein and string frames \cite{gibbons,
horowitz, gurses, cai} in Riemannian geometries. We emphasize
again that classical solutions of the coupled field equations
given in the Brans-Dicke, Einstein and string frames, whether we
consider a space-time geometry with or without torsion, are  all
conformally equivalent to each other. However, the scale invariant
case can be most conveniently studied in the Brans-Dicke frame
\cite{dereli2}. In terms of  spherical polar coordinates $ ( t,r,
\theta_{i}, i=1,2,3, \cdots ,D-2 ) $, we take the metric
\begin{equation}
g=-f^{2}(r)dt\otimes dt+h^{2}(r)dr\otimes dr+R^{2}(r)d\Omega _{D-2} ,
\end{equation}
axion field $(D-2)$-form
\begin{equation}
H = g(r) e^{1} \wedge e^{2} \wedge e^{3} \wedge \cdots  \wedge e^{D-2} ,
\end{equation}
and the dilaton scalar
\begin{equation}
\phi = \phi ( r ).
\end{equation}

\noindent {\bf Case:} $c \not= 0$, $k \not= - \frac{D-4}{D-2}$.

\bigskip

The solutions are given by the metric functions \cite{dereli2}
\begin{equation}
\begin{array}{c}
R(r)=r\left( 1-\left( \frac{C_{1}}{r}\right) ^{n-2}\right) ^{\alpha _{3}} \\
f(r)=\left( 1-\left( \frac{C_{2}}{r}\right) ^{n-2}\right) ^{\alpha
_{4}}\left( 1-\left( \frac{C_{1}}{r}\right) ^{n-2}\right) ^{\alpha _{5}} \\
h(r)=\left( 1-\left( \frac{C_{2}}{r}\right) ^{n-2}\right) ^{\alpha
_{2}}\left( 1-\left( \frac{C_{1}}{r}\right) ^{n-2}\right) ^{\alpha _{1}}
\end{array}
\end{equation}
together with
\begin{equation}
\phi(r) = \left(1-\left(\frac{C_{1}}{r}\right)^{n-2} \right)^{\frac{2
\gamma}{\alpha}}
\end{equation}
and
\begin{equation}
g(r)=\frac{Q}{R^{D-2}}
\end{equation}
where the exponents are related by
$$
\alpha_{1}=\gamma\left(\frac{1}{(D-3)}-\frac{2}{(D-2)\alpha}\right)-\frac{1}{2}
\quad , \quad \alpha_{2}=-\frac{1}{2} \qquad ,
$$
$$
\alpha_{3}=\left(\frac{1}{(D-3)}-\frac{2}{(D-2)\alpha}\right)\gamma
\quad ,
$$
$$
\alpha_{4}=\frac{1}{2}  \quad , \quad
\alpha_{5}=\frac{1}{2}-\left(1+\frac{2}{(D-2)\alpha}\right)\gamma .
$$
Here, we introduced  parameters
\begin{equation}
\gamma = \frac{ (D-2) \alpha^{2} }{4 c (D-3) + (D-2) \alpha^{2} }
\end{equation}
and
\begin{equation}
c = \omega + \frac{ D-1 }{D-2} .
\end{equation}
The integration constants $C_{1}$ and $C_{2}$ should satisfy
\begin{equation}
Q^{2} = \frac{4 c ( C_{1} C_{2} )^{D-3} (D-3)^{2} }{ \alpha^{2} } .
\end{equation}
These solutions are asymptotically flat as $ r \rightarrow \infty $. Therefore
the following physical constants can be identified:

\noindent We define the mass
\begin{equation}
2M \equiv \lim_{r \rightarrow \infty} r^{D-3}
(1-f^{2})= (C_{2})^{D-3} + \left(1-\frac{4 \gamma }{(D-2) \alpha }-2\gamma
\right)(C_{1})^{D-3}
\end{equation}
The scalar charge is
\begin{equation}
\Sigma \equiv \lim_{r \rightarrow \infty} \frac{\phi^{\prime}}{\phi}
r^{D-2} =
2(D-3) (C_{1})^{D-3} \frac{\gamma}{\alpha} .
\end{equation}
Magnetic charge can be found from
\begin{equation}
Q \equiv \lim_{r \rightarrow \infty } g r^{D-2} = Q .
\end{equation}
Eliminating the integration constants $(C_{1})^{D-3}$ and
$(C_{2})^{D-3}$ above, we obtain one relationship among our three
physical parameters:
\begin{equation}
Q^{2} = \frac{2(D-3) \Sigma }{ \alpha } c \left\{ \left(2 \gamma + \frac{4 \gamma }{(D-2) \alpha } -1 \right) \frac{ \Sigma \alpha }{2(D-3) \gamma } + 2 M \right\}
\end{equation}
The BPS bound is determined from this relationship  since $\Sigma$
is a real parameter. This implies the following inequality
satisfied by the mass and charge:
\begin{equation}
M \geq \sqrt{ \frac{4 c (D-3) - 4 \alpha - (D-2) \alpha^{2} }{4 c (D-2) (D-3)^{2} } }|Q|
\end{equation}
provided
\begin{equation}
\alpha^{2}(D-2) + 4 \alpha \leq 4 c (D-3) .
\label{27}
\end{equation}
Assume that $C_{2} > C_{1} $. Then, for $Q \not= 0 $, the metric functions admit an
outer horizon at $ r_{+} = C_{2} $ and an
inner horizon at $ r_{-} = C_{1} $. The corresponding curvature scalar
(of the Levi-Civita connection)
\begin{eqnarray}
^{(0)}{\cal R} = \frac{1}{r^{2(D-2)}} \{ \left( \frac{D-4}{D-2} - \frac{(D-1)
\alpha}{ c (D-2)} \right) Q^{2} \left(1-\left(\frac{C_{1}}{r}
\right)^{D-3} \right)^{ \{\frac{2(k-1) \gamma}{\alpha} -
2(D-2) \alpha_{3} \}} \nonumber \\
- \omega \left( \frac{2 \gamma}{\alpha} (C_{1})^{D-3} (D-3) \right)^{2}
\left(1-\left(\frac{C_{1}}{r} \right)^{D-3} \right)^{-2-2 \alpha_{1}}
\left(1- \left(\frac{C_{2}}{r} \right)^{D-3} \right) \}
\end{eqnarray}
is finite at $ r_{+} = C_{2} $. The calculation of quadratic curvature invariant on the other hand yields
\begin{equation}
\ast ( R_{ab} \wedge \ast R^{ab} ) \sim  \left( 1- \left( \frac{C_{1}}{r} \right)^{D-3} \right)^{-4-4 \alpha_{1} } r^{-4(D-2)} ,
\end{equation}
which shows that $r=0$ is an essential singularity. So the solutions above
describe black holes.

It is also interesting to see that if geometry of space-time is equipped with a
connection with torsion, then the corresponding curvature scalar
\begin{eqnarray}
{\cal R} = \frac{1}{r^{2(D-2)}}\{ \left(\frac{D-4}{D-3} \right) Q^{2} \left(1 - \left( \frac{C_{1}}{r} \right)^{D-3} \right)^{ \{ \frac{2(k-1) \gamma }{ \alpha }-2(D-2) \alpha_{3} \}}  \nonumber \\
- c \left( \frac{2 \gamma }{ \alpha } (C_{1})^{D-3} (D-3) \right)^{2}
 \left(1- \left(\frac{C_{1}}{r} \right)^{D-3} \right)^{ -2-2 \alpha_{1} }
\left(1- \left( \frac{C_{2}}{r} \right)^{D-3} \right) \}
\end{eqnarray}
is again finite at $ r_{+} = C_{2} $ while $ r=0 $ is an
essential singularity. Hence, the nature of the horizon and
the essential singularity are not affected by torsion.

\bigskip

\noindent {\bf Case:} $c=0$, $k=- \frac{D-4}{D-2} $.

\bigskip

A class of asymptotically flat solutions to conformally scale
invariant theory has the following form:
\begin{equation}
\begin{array}{c}
R(r)=r \left(1-\left(\frac{E_{1}}{r} \right)^{D-3} \right)^{-\frac{\beta}{(D-2)}} \\
f(r)=\left(1- \left(\frac{E_{1}}{r}\right)^{D-3} \right)^{\frac{1}{2} -
\frac{\beta}{(D-2)}} \left(1-\left(\frac{E_{2}}{r} \right)^{D-3} \right)^{1/2} \\
h(r)=\left(1-\left(\frac{E_{1}}{r} \right)^{D-3} \right)^{-\frac{1}{2}-\frac{\beta}{(D-2)} } \left(1-\left(\frac{E_{2}}{r} \right)^{D-3} \right)^{-\frac{1}{2}} \\
\phi(r)=\left(1-\left(\frac{E_{1}}{r}\right)^{D-3} \right)^{\beta} \\
g(r)=\frac{Q}{R^{D-2}}
\end{array}
\label{con}
\end{equation}
where $ E_{1} $ and $E_{2}$ are constants that satisfy,
\begin{equation}
(E_{2}E_{1})^{D-3} = \frac{Q^{2}}{(D-2)(D-3)} .
\end{equation}
$\beta$ is a free parameter.
The special case of parameter values $Q=0$ and $E_{2}=0$ brings (\ref{con}) to
Einstein-conformal scalar field solution of Bekenstein \cite{bekenstein1}.
Bekenstein proposed a black hole interpretation of his solutions based
on the study of conformal world lines \cite{bekenstein2}. The scalar particles are
postulated to follow geodesic world-lines in Brans-Dicke theory.
On the other hand, if space-time geometry is equipped with a
connection with torsion, history of particles would be an autoparallel of a
connection with torsion \cite{dereli4}. It has been shown that the conformal world-lines are
nothing but the autoparallel curves in the non-Riemannian reformulation
of the Brans-Dicke theory \cite{dereli3}. In this case, the scalar curvature
of the connection with torsion is calculated as
\begin{equation}
{\cal R}_{c} = \frac{D-4}{D-2} Q^{2} \left( 1 - \left( \frac{E_{1}}{r} \right)^{D-3} \right)^{\frac{2 \beta }{(D-2)} } \frac{1}{r^{2(D-2)}} .
\end{equation}
It is seen that $r=E_{2}$ is a regular event horizon, while $r=0$ is an
essential singularity. Therefore conformal solutions describe a
black hole. Mass of the black hole can be defined in terms of integration constants:
\begin{equation}
2 M_{c} = (E_{2})^{D-3} + \left( 1 - \frac{2 \beta }{(D-2)} \right) (E_{1})^{D-3} .
\end{equation}
$ \beta $ turns out to be proportional to the scalar charge.

\section{Conclusion}

In this paper we have studied axi-dilaton gravity theories in $D
\geq 4$ dimensional space-times. We have shown by making use of
the conformal rescaling properties of the space-time geometry, the
equivalence of the variational field equations obtained in the
Brans-Dicke, Einstein and string frames, with or without torsion.

We have investigated a class of asymptotically flat, static,
spherically symmetric solutions in the Brans-Dicke frame. The
black hole configurations found in the case of non-scale invariant
axi-dilaton gravity generalize the well-known $D=4$
Janis-Newman-Winicour solutions of the Einstein-Maxwell-massless
scalar field equations \cite{janis}. The fact that we are working
in the Brans-Dicke frame is essential to our discussion of the
solutions of the scale invariant axi-dilaton gravity in
$D$-dimensions. The solutions found in this case generalize the
 conformal black hole solutions of Bekenstein of $D=4$
Einstein-conformal scalar field theory.

\end{document}